\documentstyle[12pt]{article}
\renewcommand
\baselinestretch{1.4}

\def\brr{\begin{array}}
\def\beq{\begin{equation}}
\def\ben{\begin{enumerate}}
\def\een{\end{enumerate}}
\def\err{\end{array}}
\def\eeq{\end{equation}}
\def\bea{\begin{eqnarray}}
\def\eea{\end{eqnarray}}

\def\nn{\nonumber}

\begin{document}



\hfill March 1993

\vspace*{3mm}

\begin{center}

{\LARGE \bf
2D dilaton-Maxwell gravity as a fixed point of the renormalization
group}
\vspace{4mm}

\renewcommand
\baselinestretch{0.8}

{\sc E. Elizalde}\footnote{E-mail address: eli @ ebubecm1.bitnet}
\\
{\it Department E.C.M., Faculty of Physics, University of
Barcelona, \\
Diagonal 647, 08028 Barcelona, Spain} \\
{\sc S. Naftulin} \\
{\it Institute for Single Crystals, 60 Lenin Ave., 310141 Kharkov,
Ukraine} \\  and \\
{\sc S.D. Odintsov}\footnote{ E-mail address:
odintsov @ theo.phys.sci.hiroshima-u.ac.jp} \\ {\it
Tomsk Pedagogical Institute, 634041 Tomsk, Russia,} \\ and {\it
Department of
Physics, Faculty of Science, Hiroshima University, \\
Higashi-Hiroshima 724, Japan}
\medskip

\renewcommand
\baselinestretch{1.4}

\vspace{5mm}

{\bf Abstract}
\end{center}

A general model of dialton-Maxwell gravity in two dimensions is
investigated. The corresponding one-loop effective action and the
generalized $\beta$-functions are obtained. A set of models that
are fixed points of the renormalization group equations are
presented.

\vspace{3mm}


\newpage

Recently, the issue of the Hawking evaporation of black holes [1]
has been discussed in two-dimensional (2D) dilaton gravity with
matter, what is considered to be a toy model for 2D gravity. Based
on ref. [2], different aspects for different models of 2D quantum
dilaton gravity ---including relevant questions about the
associated black holes--- have been extensively considered in the
recent literature (for a review and a quite comprehensive list of
references see, for example, [3]). Due to the wide variety of
models of 2D gravity under current investigation (some of which are
classically equivalent) the following question arises naturally: what
is a good criterion in order to select reasonable models of 2D
gravity?

In the present letter we suggest one: to use the renormalization
group as such a criterion ---in a sense that will become clear in
the progress of this work.
We start from the general model of 2D gravity defined by the action
\begin{equation}
S=-\int d^{2}x \,\sqrt{g}\left[ {1\over 2} Z( \Phi ) g^{\mu
\nu}\partial_\mu \Phi  \partial_\nu \Phi + C(\Phi ) R +V(\Phi ) +
{1\over 4} f( \Phi ) F_{\mu \nu}^2 \right],
\label{so}
\end{equation}
where $\Phi$ is the dilaton field,  $Z(\Phi )$, $ C(\Phi )$, $V
(\Phi )$ and  $f(\Phi )$ are arbitrary functions, and $F_{\mu \nu}
= \partial_\mu A_\nu-  \partial_\nu A_\mu$. Notice that the model
(1) is related, via some compactification, with the
 four-dimensional Einstein-Maxwell theory, which admits charged
black hole solutions [4]. For some specific values of the functions
in (1), this general model corresponds, in fact, to the heterotic
string efective action [5,6]. Notice also that the theory (1) is
multiplicatively renormalizable [8,9].

The one-loop effective action for theory (1) with the choice:
$Z (\Phi )=1$, $C (\Phi )=$ const.,  and  $V (\Phi )$ and  $f(\Phi
)$ arbitrary, has been calculated in different gauges in ref. [7].
Here, we are going to find the one-loop effective action using
the background field formalism for this model (1),
 in the general case.

In the background field method, we split
\begin{equation}
g_{\mu\nu} \longrightarrow \bar{g}_{\mu\nu} =g_{\mu\nu}
+h_{\mu\nu}, \ \ \ \ \
\Phi \longrightarrow \bar{\Phi} = \Phi +
\varphi,  \ \ \ \ \
A_\mu \longrightarrow \bar{A}_mu =A_\mu +Q_\mu,
\end{equation}
where $h_{\mu\nu}$, $\varphi$ and $Q_\mu$ are the quantum fields.
The simplest minimal gauge fixing action is given by
\beq
S_{GF}= -{1\over 2}\int d^2 x\,  \chi^A c_{AB} \chi^B,
\eeq
where $A \equiv \{ \mu, * \}$, $c_{\mu\nu} =-C\sqrt{ g} \,
g_{\mu\nu}$,
 $ \chi^\mu =-\nabla_\nu \bar{h}^{\mu\nu} +\frac{C'}{C} \nabla^\mu
\varphi$, $C_*=\sqrt g f$, $\chi_* =-\nabla^\nu Q_\nu$.
Working in this gauge we obtain the following effective action
(details of the very involved calculations will be reported
elsewhere):
\begin{eqnarray}
\Gamma_{div}=-{1\over 2\epsilon}\int
d^2x\,\sqrt{g}\,\Biggl\{5R+{2\over C}V
             +{2\over
C'}V'+\left({f'\over2C'}-{f\over2C}\right)F^2\qquad\cr\cr
+\left({f'\over f}+{C'\over C}-{Z\over C'}\right)(\Delta\Phi)
\quad\cr\cr
+\left({f''\over f}-{{f'^2}\over f^2}+{C''\over C}-3{{C'}^2\over
C^2}-
                 {C''Z\over{C'}^2}\right)(\nabla^\lambda\Phi
)(\nabla_\lambda
                 \Phi) \Biggr\} \ .
\label{ea1}
\end{eqnarray}
It must be remarked that in this expression all
surface divergent terms have been kept.

Let us  discuss the on-shell limit of the effective action
(\ref{ea1}). The corresponding classical field equations are
\bea
\frac{\delta S}{\delta \Phi} &=& - \nabla_\nu
(Zg^{\mu\nu}\partial_\mu \Phi ) + \frac{1}{2} Z'(\nabla^\mu\Phi
)(\nabla_\mu\Phi) +C' R + V' + \frac{1}{4} f' F^2 =0, \nn \\
g^{\mu\nu}\frac{\delta S}{\delta g^{\mu\nu}} &=& -\Delta C +V -
\frac{1}{4} f F^2 =0,
\label{ea2}
\eea
and substituting eqs. (\ref{ea2}) into the effective action
(\ref{ea1}), keeping
all the surface counterterms, we obtain
\beq
\Gamma_{div}^{on-shell}=-{1\over 2\epsilon}\int d^2 x\,\sqrt g\,
\Biggl\{ 3R+ \Delta \left[ \ln (fC^3) \right] + \nabla^\lambda
\left[\frac{Z}{C'} \nabla_\lambda (\Phi ) \right]\Biggr\}.
\label{}
\eeq
As we see, the theory considered is finite on-shell.

We will now study the renormalization structure and
renormalization group equations for dilaton gravity with the action
(1). One could, of course, discuss the renormalization of the metric and find
th
restrictions imposed by off-shell multiplicative
renormalizability  (in the usual sense) on the form of the
functions  $Z(\Phi )$, $ C(\Phi )$,  $V (\Phi )$ and $f (\Phi )$.
However, instead of
doing this, we will not renormalize the fields but rather will
consider the functions $Z$, $C$, $V$ and $f$ as generalized
effective
couplings (generalized renormalizability). Then the generalized
$\beta$-functions can be found and generalized renormalization
group equations are generated.

 The general structure of renormalization for general
couplings is given by
\beq
Z_0 = (\mu^2)^{\epsilon'} \left[ Z+ \sum_{k=1}^\infty \frac{a_{kZ}
(Z,C,V,f)}{{\epsilon'}^k} \right],
\eeq
where $\epsilon'=n-2$, and similar expressions for $C$, $V$ and
$f$. As
it follows from one-loop renormalization, eq. (\ref{ea1}),
\beq
a_{1Z}=-\frac{Z'}{C'}+\frac{2{C'}^2}{C^2}+\frac{2C''Z}{{C'}^2}, \
\ \
 a_{1C}=0,  \ \ \
 a_{1V}=-\frac{V}{C}-\frac{V'}{C'}, \ \ \
 a_{1f}=\frac{f}{C}-\frac{f'}{C'}.
\label{aes}
\eeq
Now, the generalized one-loop $\beta$-functions are given by the
standard relations:
\beq
\beta_T = -a_{1T} + Z \frac{\delta a_{1T}}{\delta Z} + C
\frac{\delta a_{1T}}{\delta C} +
 V \frac{\delta a_{1T}}{\delta V}+
 f \frac{\delta a_{1T}}{\delta f},
\label{bts}
\eeq
where $T\equiv \{ Z,C,V,f \}$. Using (\ref{aes}) and (\ref{bts}) we obtain
\bea
\beta_C &=& 0, \nn \\
\beta_V &=& \frac{V}{C}+\frac{V'}{C'}- \frac{V C''}{{C'}^2}-
\frac{CV''}{{C'}^2}+2\frac{CV'C''}{{C'}^3} \nn \\
\beta_Z &=& \frac{Z'}{C'}+\frac{2{C'}^2}{C^2} -
\frac{4C''}{C}-\frac{ZC''}{{C'}^2}+3 \frac{CZ''}{{C'}^2}-
2\frac{CZ'C''}{{C'}^3},  \\
\beta_f &=& -\frac{f}{C}+\frac{f'}{C'}- \frac{fC''}{{C'}^2}
- \frac{Cf''}{{C'}^2}+2 \frac{Cf'C''}{{C'}^3}.  \nn
\label{bes}
\eea

The renormalization group equations have the following form:
\beq
\frac{\partial C}{\partial t} = \beta_C, \ \ \ \ \ \ \
\frac{\partial
V}{\partial t} = \beta_V, \ \ \ \ \ \ \
\frac{\partial Z}{\partial t} = \beta_Z, \ \ \ \ \ \ \
\frac{\partial f}{\partial t} = \beta_f,
\label{dts}
\eeq
with $t=\ln \mu^2$. The system of equations (\ref{dts}) is very
difficult
to solve. It depends not only on the scaling parameter $t$, as in
usual field theory, but also on some unknown functions of the field
variables. Moreover, nobody has any idea about the proper boundary
conditions (initial data) that the partial differential equation
system (\ref{dts}) should satisfy.

Notwithstanding that, we can extract some useful information yet from
the renormalization group equations (\ref{dts}). In particular, we can
look for
fixed points of this system, what does not at all involve the
knowledge of initial data. The equations they must satisfy are
\beq
\beta_C=0, \ \ \ \ \ \ \ \beta_V=0,
\ \ \ \ \ \ \ \beta_Z=0,
\ \ \ \ \ \ \ \beta_f=0.
\label{bos}
\eeq
The system of differential equations (\ref{bos}) is still very
complicated. Nevertheless, some basic, particular solutions of the
same can be discovered. For example, motivated by the CGHS action
[2], we can look for fixed points of the following type:
\beq
C=e^{a_1\Phi}, \ \ \ \ V=e^{a_2\Phi},
 \ \ \ \ Z=e^{a_3\Phi},
 \ \ \ \ f=e^{a_4\Phi},
\label{exps}
\eeq
where $a_1$, $a_2$, $a_3$ and $a_4$ are some constants. Substituting
(\ref{exps}) into eq. (\ref{bos}) we obtain the following equations:
\bea
&&a_1= \frac{a_2(a_2-1)}{a_2+1}, \nn \\
&&3a_3^3+a_3(1-3a_1)+2a_1(1-2a_1) =0, \\
&&a_4^2-3a_1a_4+2a_1^2 =0, \nn
\eea
which have the solution
\bea
&& a_1=0,  \ \ \ a_2=1, \ \ \ a_3 =0,- \frac{1}{3}, \ \ \ a_4=0; \nn \\
&& a_1=\frac{2}{3}, \ \ \ a_2=2, \ \ \ a_3 =\frac{1}{6} \left( 1\pm
\sqrt{19/3} \right), \ \ \ a_4= \frac{2}{3}, \frac{4}{3}; \ \ \ \mbox{etc.}
\eea
In the same way, different particular cases of fixed points can be
considered. For instance, for
\beq
C= \Phi^{\alpha_1}, \ \ \ \ V= \Phi^{\alpha_2},
 \ \ \ \ Z= \Phi^{\alpha_3},
 \ \ \ \ f= \Phi^{\alpha_4},
\eeq
we get
\bea
&& \alpha_1 (\alpha_1 -2)=0, \nn \\
&& \alpha_2^2+ (1-3\alpha_1)\alpha_2 -\alpha_1 =0, \nn \\
&& 3\alpha_3^2-(1+\alpha_1)\alpha_3-\alpha_1(\alpha_1-1)=0, \\
&& \alpha_4^2+(1-3\alpha_1)\alpha_4+\alpha_1(2\alpha_1-1)=0. \nn
\eea
Particular solutions are
\bea
&&\alpha_1=0, \ \ \ \alpha_2=-1,  \ \ \ \alpha_3 =\frac{1}{3},
\ \ \ \alpha_4=-1; \nn \\
&&\alpha_1=2, \ \ \ \alpha_2=\frac{1}{2}
\left(5\pm \sqrt{33} \right),
\ \ \ \alpha_3=\frac{1}{2} \left(5\pm \sqrt{11/3} \right),
\ \ \ \alpha_4=2, 3; \ \ \
 \ \ \ \mbox{etc.}
\eea
Guided by the CGHS model, one could also try to find 'mixed'
 solutions of the type
\beq
C= \Phi^{a_1} e^{b_1\Phi}, \ \ \ \
V= \Phi^{a_2} e^{b_2\Phi}, \ \ \ \
Z= \Phi^{a_3} e^{b_3\Phi}, \ \ \ \
f= \Phi^{a_4} e^{b_4\Phi}.
\eeq
It can be shown, however, that no
 new solution is found in this way: the
only possibilities reduce to the cases above (namely, either all $a_i$
or all $b_i$ vanish).

Thus, we have shown that (at least in principle) one can find
ultraviolet stable fixed points for the generalized couplings
$Z(\Phi )$, $ C(\Phi )$ $V(\Phi )$ and $f (\Phi )$.

Summing up, we have here investigated
the one-loop renormalization structure of a general model of  2D
dilaton-Maxwell gravity (1), and the
 divergences of the corresponding one-loop effective action have been
found.
Generalized renormalizability of the general model of dilaton
gravity has been discussed, and the corresponding generalized
$\beta$ functions have been obtained.

 The analysis of the
renormalization group equations yields some set of generalized
couplings $\{ Z,C,V,f \}$ which are fixed points of such equations.
It is interesting to notice that the CGHS model does {\it not}
belong to this set. A further remark is the fact that if one
requires the renormalizability of the theory in the usual sense,
then one can renormalize the metric of the theory (1) through the
following transformation
\beq
g_{\mu\nu} = \exp \left[ \frac{1}{\epsilon} \left( \frac{1}{C}+
\frac{Z}{2{C'}^2} \right) \right] g_{\mu\nu}^R.
\eeq
It turns out that the 2D dilaton-Maxwell theory (1)
 is multiplicatively
renormalizable off-shell in the usual sense for the following
choice of potentials:
\beq
V= \exp \left[ aC + \int \frac{Z\, d\Phi}{2C'}\right], \ \ \ \ \
f= \exp \left[-bC - \int \frac{Z\, d\Phi}{2C'} \right],
\eeq
where $a$ and $b$ are arbitrary constants. For the theory of
ref. [7] the
above potentials are of Liouville type, what is in full agreement
with ref. [7]. We observe, finally, that the condition of
off-shell renormalizability in the usual sense can be considered
also as a good criterion for selecting 2D gravity models.
\vspace{5mm}

\noindent{\large \bf Acknowledgments}

We are grateful to I. Antoniadis, F. Englert,
 Y. Kazama, H. Osborn, A.A.
Slavnov and I.V. Tyutin for useful discussions at different stages
of this work.
S.D.O. wishes to thank the Japan Society for the Promotion of
Science (JSPS, Japan) for financial support  and the
Particle Physics Group at Hiroshima University for kind
hospitality.
E.E. has been supported by DGICYT (Spain), research project
PB90-0022, and by the Generalitat de Catalunya.


\newpage



\begin{thebibliography}{99}

\bibitem{} S.W. Hawking, Commun. Math. Phys. {\bf 43} (1975) 199.

\bibitem{} C.G. Callan, S.B. Giddings, J.A. Harvey and A.
Strominger,  Phys. Rev. {\bf D45} (1992) 1005.

\bibitem{}  J.A. Harvey and A. Strominger, preprint EFI-92-41
(1992).

\bibitem{} G.W. Gibbons,  Nucl. Phys. {\bf B207} (1982) 337;
G.W. Gibbons and K. Maeda,  Nucl. Phys. {\bf B298} (1988) 74.

\bibitem{} C.G. Callan, D. Friedan, E. Martinec and M.J. Perry,
Nucl. Phys. {\bf B262} (1985) 593; C.G. Callan, I. Klebanov and
M.J. Perry, Nucl. Phys. {\bf B278} (1986) 78.

\bibitem{} M. McGuigan, C. Nappi and S. Yost, IAS preprint IASSNS-
HEP-91157 (1991).

\bibitem{} E. Elizalde and S.D. Odintsov, Mod. Phys. Lett. {\bf A8}
(1993) 33; E. Elizalde, S. Naftulin and S.D. Odintsov, preprint
HUPD-93-03 (1993).

\bibitem{} T. Banks and M. O'Loughlin, Nucl. Phys. {\bf B362}
(1991) 649.

\bibitem{} S.D. Odintsov and I.L. Shapiro, Phys. Lett. {\bf B263}
(1991) 183.



 \end{thebibliography}
\end{document}